\documentstyle[12pt]{article}
\def\largelinestretch{\renewcommand{\baselinestretch}{1.3}}
\largelinestretch\small\normalsize

\title{
\vspace*{-10mm}
\hfill
\parbox{4cm}{\large JINR E2-96-432\\
                    FAU-TP3-96/5}\\
\vspace*{10mm}
 SPIN-$\frac{3}{2}$ FIELDS IN THE HEAVY-BARYON EFFECTIVE THEORY}
 \author{
A.A.Bel'kov${}^{1}$,
M.Dillig${}^2$
\\
\\ \normalsize
${}^1$
        Particle Physics Laboratory,
 Joint Institute for Nuclear Research,\\ \normalsize
       RU-141980 Dubna, Moscow Region, Russia\hfill\\ \normalsize
${}^2$
Institut f\"ur Theoretische Physik III der Universit\"at Erlangen-N\"urnberg,\\
 \normalsize
D-91058 Erlangen, Germany\hfill\\
}
\date{}

\begin{document}
\thispagestyle{empty}
\begin{titlepage}
\thispagestyle{empty}
\maketitle
\begin{abstract}

    In the framework of the path integral approach we develop the
velocity component formalism for spin-${1\over2}$ and -${3\over2}$
baryons coupled with pseudoscalar mesons in the limit of heavy baryons.
    Starting from the most general chiral meson-baryon lagrangian
including the invariance of spin-${3\over2}$ fields under point
transformation we detail various problems in integrating out the
heavy velocity components in the baryon fields and derive the effective
lagrangian for the light components up to $O(1/M_{baryon})$.

\end{abstract}

\vspace{20mm}

Talk at the XIII International Seminar on High Energy Physics Problems,

Dubna, September 2-7, 1996

\end{titlepage}

    The chiral symmetry and its breaking are the fundamental principles
governing the dynamics of low-energy interactions of mesons and baryons
\cite{Weinberg}.
    The new formulation of the effective chiral lagrangians for
the meson-baryon system given in Ref.\ \cite{Jenkins} is based on
the treatment of baryons as heavy static fields with the definite velocity.
    The main advantage of this approach is that the improved derivative
expansion for the light velocity components makes it possible to include
baryons in the power counting scheme of chiral perturbative theory and
calculate the meson-baryon loops in the most systematical way.
    In such an approach some part of the higher order counterterms can
be derived using the path integral techniques as $1/M$-corrections
when the heavy velocity components of baryon fields are integrated out
of the generating functional for the meson-baryon system.

    It is well known that both free and interaction lagrangians
for a spin-$3\over2$ field coupled to a nucleon and a pseudoscalar
meson field have to be constructed in such a way that the total
lagrangian is invariant under the so called ``point transformation''
\cite{Delta,Pascalutsa}.
    This condition is a consequence of the invariance of the physical
properties of spin-$3\over2$ field with respect to rotations in the
spin-$1\over2$ space.
    The effective lagrangian introduced in the Ref.\ \cite{Jenkins}
is not invariant under point transformation and it has to be modified by
introducing additional off-shell terms which restore its point transformation
invariance.
    Such an extension also leads to the appearence of new terms when
calculating $1/M$-corrections to the effective lagrangian for the light
velocity components.

   As an extension of the formalism from Ref.\ \cite{Jenkins} the
lagrangian for the system of spin--${1\over2}$ and spin-${3\over2}$
baryons coupled with pseudoscalar mesons, which is explicitly invariant
under point transformation, is given in the form
\begin{eqnarray}
{\cal L}_{tot} &=& \overline{B}\big(
  i{{\,\not}\! D}-M-\gamma_5{{\,\not\!}{\hat A}}_\Omega \big)B
\nonumber \\ &&
 -\overline{T}^\mu\bigg(\,
       \big(i{{\,\not}\! D}-M\big)g_{\mu\nu}
      -{1\over4}\gamma_\mu \gamma_\lambda \big(i{{\,\not}\! D}-M\big)
                \gamma^\lambda \gamma_\nu
      +{\cal H}\gamma_5 \Omega_{\mu\nu}\bigg)T^\nu
\nonumber \\ &&
 +{\cal C}\Big(\, \overline{T}^\mu\Theta_{\mu\nu}A^\nu_\Omega B
               +\overline{B}A^\mu_\Omega\Theta_{\mu\nu}T^\nu
          \Big)\,,
\label{Ltot}
\end{eqnarray}
where $B$ is the spin-${1\over2}$ baryon field, while the field $T_\mu$
is related with the vector-spinor representation $\psi_\mu$ for
spin-${3\over2}$ field, introduced by Rarita and Schwinger
\cite{Rarita}
\begin{equation}
T_\mu = {\cal O}^A_{\mu\nu}\psi^\nu\,,\quad
{\cal O}^A_{\mu\nu} = g_{\mu\nu}+{1\over2}A\gamma_\mu\gamma_\nu\,,
\label{Tfield}
\end{equation}
with
\begin{equation}
\Theta_{\mu\nu} =  g_{\mu\nu}+z\gamma_\mu \gamma_\nu\,,
\label{Theta}
\end{equation}
where $z$ and $A$ are arbitrary parameters with $A \neq -{1\over2}$.
   In Eq.\ (\ref{Ltot}) the covariant derivative
$D^\mu{*} =\partial^\mu{*}+\big[ V^\mu_\Omega ,{*}\big]$ couples the baryon
fields with the vector combination of the pseudoscalar meson matrix
$\Omega$ and its derivative,
$$
V_\Omega^\mu = {1\over2}\big( \Omega\partial^\mu\Omega^\dagger
                             +\Omega^\dagger \partial^\mu\Omega\big)\,,
$$
while the operators
\begin{equation}
{\hat A}_\Omega^\mu = D\big\{A_\Omega^\mu,B\big\}
                     +F\big[ A_\Omega^\mu,B\big]\,,
\label{Aoper}
\end{equation}
\begin{equation}
\Omega_{\mu\nu} =  {{\,\not}\! A}_\Omega g_{\mu\nu}
      +g_1\big(\gamma_\mu A_{\Omega\nu}+A_{\Omega\mu}\gamma_\nu\big)
      +g_2\gamma_\mu {{\,\not}\! A}_\Omega \gamma_\nu\,,
\label{Omega}
\end{equation}
couple baryons with the axial-vector combination
$$
A_\Omega^\mu = {i\over2}\big( \Omega\partial^\mu\Omega^\dagger
                             -\Omega^\dagger \partial^\mu\Omega\big)\,.
$$
   In nonlinear realization of chiral symmetry in $SU(3)$ the matrix
$\Omega$ for pseudoscalar mesons is given by
$$
   \Omega =\mbox{exp}\bigg(\frac{i}{\sqrt{2}F_0}
           \sum^{8}_{a=1}\varphi^a\frac{\lambda^a}{2}\bigg)\,,
$$
where $\varphi^a$ are the pseudoscalar octet fields, and $F_0$ is
the bare $\pi$ decay constant.
    For simplicity the flavour $SU(3)$ indices in the lagrangian
(\ref{Ltot}) are omitted, and $M$ being an avaraged $N \Delta$ mass.

   As recently stressed by Pascalutsa \cite{Pascalutsa}, the
representation (\ref{Ltot}) is convenient for the description of
processes with spin-${3\over2}$ fields off-shell, as the full
$A$-dependence is hidden now in the new fields (\ref{Tfield})
which are explicitly invariant under point transformation
\begin{equation}
\psi_\mu \to \psi_\mu^{'} = {\cal O}^b_{\mu\nu}\psi^\nu\,,
\quad
A \to A^{'}= \frac{A-b}{1+2b}\,,
\label{point}
\end{equation}
where $b\neq -{1\over2}$ is an arbitrary parameter.
    The invariance of the lagrangian (\ref{Ltot}) under the
transformation (\ref{point}) implies that the physical content of the
theory does not depend on the choice of $A$.
    As the Rarita-Schwinger vector-spinor $\psi_\mu$ can be constructed
from the direct product of the relativistic states $\varepsilon_\mu$
for spin-1 and $u$ for spin-${1\over2}$,
$\psi_{\mu ,s}(p) =
 L^{(1)}_{\mu\nu}\big[ \varepsilon_\nu u(p)\big]_{3/2 s}$,
where the brackets denote the coupling of the Dirac spinor with the
polarization vector $\varepsilon_\mu$; $L^{(1)}(p)$ is the corresponding
boost operator, it obeys in momentum space
$$
    p_\mu \psi^\mu =0
$$
corresponding to the Lorentz condition for $\varepsilon_\mu$.
    In addition, on mass shell the supplementary condition
\begin{equation}
\gamma_\mu \psi^\mu =0
\label{subsid}
\end{equation}
is satisfied, resulting from the elimination of the spin-${1\over2}$
components from the spinor-vector direct product.
    The on-shell coupling constants $D$, $F$, ${\cal H}$ and
${\cal C}$ in Eqs.\ (\ref{Ltot}) and (\ref{Aoper}) agree with
Ref.\ \cite{Jenkins}, while $g_1$, $g_2$ and $z$ in Eqs.\
(\ref{Omega}) and (\ref{Theta}) are additional off-shell parameters.
    Due to condition (\ref{subsid}) the effective lagrangian
(\ref{Ltot}) coincides with the form presented in Ref.\ \cite{Jenkins} on
the mass shell of spin-${3\over2}$ particles.

    For meson-baryon interactions, at low energies when the momentum
transfer from meson to baryon is small, if to compare with the baryon
mass, the velocity of the baryon is conserved in the heavy mass limit.
    In this case the effective field theory can be written in terms of
baryon fields with the definite four-velocity $v$.
    After indroducing the velocity components for the baryon fields
and corresponding external sources in analogy with the heavy static
quark approach \cite{Mannel}, the generating functional corresponding to
lagrangian (\ref{Ltot}) can be written as
\begin{eqnarray}
&&
{\cal Z}(R^B_v,\rho^b_v,R^T_v,\rho^t_v)=
   \int{\cal D}(B_v,T_v,\Omega)\,\mbox{exp}\Big[\,
  i\!\int d^4x{\cal L}^{(v)}_{tot}
\nonumber \\ &&\quad\quad\quad\quad\quad\quad\quad\quad\quad\quad
               \quad\quad\quad\quad\quad\quad~~~
 +i\!\int d^4x\Big( \overline{R}^B_vB_v +\overline{B}_vR^B_v
                 +\overline{\rho}^b_vb_v +\overline{b}_v\rho^b_v
\nonumber \\ &&\quad\quad\quad\quad\quad\quad\quad\quad\quad\quad
               \quad\quad\quad\quad\quad\quad~~~~~~~
                 +\overline{R}^T_{v\mu}T_v^\mu
                 +\overline{T}_v^\mu R^T_{v\mu}
                 +\overline{\rho}^t_{v\mu}t_v^\mu
                 +\overline{t}_v^\mu \rho^t_{v\mu} \Big) \Big]\,.
\nonumber \\ &&
\label{ZBTv}
\end{eqnarray}
Here ${\cal D}(B_v,T_v,\Omega)$ is the path integral measure including the
velocity components
\begin{eqnarray}
&&
B = \mbox{e}^{-iM(v\cdot x)}(B_v +b_v)\,,\quad
B_v = \frac{1+{\not\! v}}{2}B\,,\quad
b_v = \frac{1-{\not\! v}}{2}B\,,
\nonumber \\ &&
T^\mu = \mbox{e}^{-iM(v\cdot x)}(T^\mu_v +t^\mu_v)\,,\quad
T^\mu_v = \frac{1+{\not\! v}}{2}T^\mu\,,\quad
t^\mu_v = \frac{1-{\not\! v}}{2}T^\mu\,,
\label{BTv}
\end{eqnarray}
$R^B_v$, $\rho^b_v$, $R^T_{v\mu}$ and $\rho^t_{v\mu}$ are the external
sources coupling to the velocity components (\ref{BTv}), and
\begin{eqnarray}
{\cal L}^{(v)}_{tot} &=& \overline{B}_v G B_v
 -\overline{b}_v {\cal G} b_v
 + \overline{B}_v H b_v +\overline{b}_v H B_v
\nonumber \\ &&
 -\overline{T}^\mu_v Q_{\mu\nu} T^\nu_v
 -\overline{t}^\mu_v {\cal Q}_{\mu\nu}t^\nu_v
 -\overline{T}^\mu_v R_{\mu\nu}t^\nu_v
 -\overline{t}^\mu_v R_{\nu\mu}T^\nu_v
\nonumber \\ &&
 +{\cal C}\Big[ \big( \overline{T}^\mu_v +\overline{t}^\mu_v \big)
                \Theta_{\mu\nu}A^\nu_\Omega\big(B_v +b_v\big)
               +\big(\overline{B}_v+\overline{b}_v\big)
                A^\mu_\Omega\Theta_{\mu\nu}
                \big(T^\nu_v +t^\nu_v \big)
          \Big]\,,
\label{Ltotv}
\end{eqnarray}
with
\begin{eqnarray*}
&&
  G = i(v\cdot D)-\gamma_5{{\,\not\!}{\hat A}}_\Omega\,,\quad
  {\cal G} = i(v\cdot D)+2M+\gamma_5{{\,\not\!}{\hat A}}_\Omega\,,\quad
  H = i{{\,\not}\! D}^\perp -\gamma_5{{\,\not\!}{\hat A}}_\Omega\,,
\nonumber \\ &&
  Q_{\mu\nu} =
       \,i{{\,\not}\! D}^\parallel g_{\mu\nu}
      -{1\over4}\gamma_\mu \gamma_\lambda
       i{{\,\not}\! D}^\parallel\gamma^\lambda\gamma_\nu
      +{\cal H}\gamma_5\Omega_{\mu\nu}\,,
\nonumber \\ &&
  {\cal Q}_{\mu\nu} =
       \, \big[ i{{\,\not}\! D}^\parallel-2M\big]g_{\mu\nu}
      -{1\over4}\gamma_\mu \gamma_\lambda
       \big[ i{{\,\not}\! D}^\parallel-2M\big]\gamma^\lambda\gamma_\nu
      +{\cal H}\gamma_5\Omega_{\mu\nu}\,,
\nonumber \\ &&
  R_{\mu\nu} =\, i{{\,\not}\! D}^\perp g_{\mu\nu}
      -{1\over4}\gamma_\mu \gamma_\lambda i{{\,\not}\! D}^\perp
                \gamma^\lambda\gamma_\nu
      +{\cal H}\gamma_5\Omega_{\mu\nu}\,,
\end{eqnarray*}
where $D_\mu^\parallel =v_\mu (v\cdot D)$,
      $D^{\perp}_\mu = D_\mu-v_\mu (v\cdot D)$.

    In order to derive Eq.\ (\ref{Ltotv}) as the effective action for
the velocity components (\ref{BTv}), we have used the gauge condition
$v_\mu T^\mu =0$, which is equivalent to this $A=0$ in (\ref{Tfield}).
    We are allowed to use this gauge condition as the physical
properties of the spin-${3\over2}$ field are independent on the
choice of parameter $A$ due to its invariance with respect to
point transformation.

    In Eqs.\ (\ref{ZBTv}) and (\ref{Ltotv}) the light components $B_v$
and $T^\mu_v$ correspond to massless effective baryon fields, while the
heavy components $b_v$ and $t^\mu_v$ have effective mass $2M$.
    As a consequence, they can be integrated from the generating
functional (\ref{ZBTv}) as the heavy degrees of freedom  performing the
Gaussian integration.
    According to the standard procedure, the heavy components
$\overline{b}_v$, $\overline{t}^\mu_v$, $b_v$ and $t^\mu_v$ have to be
replaced in the exponent of the Eq.\ (\ref{ZBTv}) by the solutions of the
equations
\begin{equation}
\frac{\delta \big( {\cal L}^{(v)}_{tot}
                  +\overline{\rho}^b_v b_v\big)}{\delta b_v}=0\,,\quad
\frac{\delta \big( {\cal L}^{(v)}_{tot}
                  +\overline{\rho}^t_{v\mu} t^\mu_v \big)}
     {\delta t^\mu_v}=0\,,
\label{vareq1}
\end{equation}
\begin{equation}
\frac{\delta \big( {\cal L}^{(v)}_{tot}
                  +\overline{b}_v \rho^b_v\big)}{\delta \overline{b}_v}
=0\,,\quad
\frac{\delta \big( {\cal L}^{(v)}_{tot}
                  +\overline{t}^\mu_v \rho^t_{v\mu} \big)}
     {\overline{t}^\mu_v}=0\,,
\label{vareq2}
\end{equation}
which leads to the system of the equations for $\overline{b}_v$ and
$\overline{t}^\mu_v$:
\begin{eqnarray}
&&
  - \overline{b}_v {\cal G}
  + {\cal C}\,\overline{t}^\mu_v\Theta_{\mu\nu}A^\nu_\Omega
  + \overline{B}_v H
  + {\cal C}\,\overline{T}^\mu_v\Theta_{\mu\nu}A^\nu_\Omega
  + \overline{\rho}^b_v = 0\,,
\nonumber \\ &&
    {\cal C}\,\overline{b}_v A^\mu_\Omega \Theta_{\mu\nu}
  - \overline{t}^\mu_v {\cal Q}_{\mu\nu}
  + {\cal C}\,\overline{B}_v A^\mu_\Omega \Theta_{\mu\nu}
  - \overline{T}^\mu_v R_{\mu\nu}
  + \overline{\rho}^t_{v\nu} = 0\,,
\label{equtb}
\end{eqnarray}
and similarly for $b_v$ and $t^\mu_v$.

    We will calculate the effective meson-baryon action in terms of
light velocity components of baryon fields up to and including
corrections of $O(1/M)$.
    In such an approximation it is enough to keep only terms up
to $O(1/M^3)$ in the solutions of the equations (\ref{equtb}):
\begin{eqnarray*}
\overline{b}_v &=&\Big(
    \overline{B}_v H
  + {\cal C}\,\overline{T}^\mu_v
    \Theta_{\mu\nu}A^\nu_\Omega
  + \overline{\rho}^b_v \Big) {\cal G}^{-1}
\nonumber \\ &&
  + \frac{{\cal C}}{4M^2}\Big(
    {\cal C}\,\overline{B}_v A_\Omega^\alpha
    \Theta_{\alpha\beta}
  - \overline{T}^\alpha_v R_{\alpha\beta}
  + \overline{\rho}^t_{v\beta} \Big)
    \widetilde{\cal Q}^{\beta\mu}\Theta_{\mu\nu} A_\Omega^\nu
  + O\bigg(\frac{1}{M^3}\bigg)\,,
\nonumber \\
\overline{t}^\mu_v &=& \Big(
    {\cal C}\,\overline{B}_v A_{\Omega\alpha} \Theta^{\alpha\nu}
  - \overline{T}_{v\alpha} R^{\alpha\nu}
  + \overline{\rho}^{t\nu}_v\Big)
    {\cal Q}^{-1}_{\nu\mu}
\nonumber \\ &&
  + \frac{{\cal C}}{4M^2}\Big(
    \overline{B}_v H
  + {\cal C}\,\overline{T}^\alpha_v
    \Theta_{\alpha\beta}A^\beta_\Omega
  + \overline{\rho}^b_v \Big)
    A_\Omega^\tau \Theta_{\tau\nu} \widetilde{\cal Q}^{\nu\mu}
  + O\bigg(\frac{1}{M^3}\bigg)\,.
\label{eqtb}
\end{eqnarray*}
    In solving for the heavy velocity components in the preceeding
equation, we stress the appearance of the inverse operators
${\cal G}^{-1}$ and ${\cal Q}^{-1}_{\mu\nu}$
given explicitly as
\begin{eqnarray*}
{\cal G}^{-1} &=& \frac{1}{2M}\bigg[ 1- \frac{1}{2M}\Big(
           i(v\cdot D)+\gamma_5{{\,\not\!}{\hat A}}_\Omega\Big)
                       \bigg]\,,
\nonumber \\
{\cal Q}^{-1}_{\mu\nu} &=& \frac{1}{2M}\bigg[
  - g_{\mu\nu} + \frac{1}{3}\gamma_\mu\gamma_\nu
  - \frac{1}{2M}\bigg(
    {{\,\not}\! D}^\parallel g_{\mu\nu}
  - \frac{1}{6}\gamma_\mu {{\,\not}\! D}^\parallel\gamma_\nu
  - \frac{1}{3}\Big\{\gamma_\mu\gamma_\nu ,
                     {{\,\not}\! D}^\parallel\Big\}
                 \bigg)
\nonumber \\ &&
  - \frac{{\cal H}}{2M}\gamma_5 \bigg(
    {{\,\not}\! A}_\Omega g_{\mu\nu}
  - \frac{2(1-g_1)-9g_2}{9}\gamma_\mu {{\,\not}\! A}_\Omega\gamma_\nu
  - \frac{1}{3}\big\{\gamma_\mu\gamma_\nu ,{{\,\not}\! A}_\Omega \big\}
\nonumber \\ &&\quad\quad\quad
  - \frac{g_1}{3}\big( \gamma_\mu A_{\Omega\nu}
                             +A_{\Omega\mu}\gamma_\nu \big)
                     \bigg) \bigg]\,
\nonumber \\ &&
    \equiv \frac{1}{2M}\widetilde{\cal Q}_{\mu\nu}+O(1/M^2)\,,
\end{eqnarray*}
with
$$
   \widetilde{\cal Q}_{\mu\nu} = -g_{\mu\nu}
                            +\frac{1}{3}\gamma_\mu\gamma_\nu\,,
$$
which obey the relations
$$
  {\cal G}{\cal G}^{-1}=1+O(1/M^2)\,,\quad
  {\cal Q}_\mu{}^\alpha{\cal Q}^{-1}_{\alpha\nu}=
  g_{\mu\nu}+O(1/M^2)\,.
$$
    Note that -- consistent with the expansion of the effective
lagrangian to $O(1/M)$ given below -- there are {\it no} corrections
of this order both in the inverse operators and the heavy velocity
components.
    The solutions for $b_v$ and $t^\mu_v$ can be obtained from the
Eq.\ (\ref{vareq2}) in analogy.

    After integrating out the heavy velocity components of
spin--${1\over2}$ and spin-${3\over2}$ baryons, the generating
functional contains only the light components $B_v$ and $T^\mu_v$:
\begin{eqnarray}
{\cal Z}&=&
   {\cal N}^{-1}\int\widetilde{{\cal D}}(B_v,T_v,\Omega)\,
   \mbox{exp}\bigg\{\, i\!\int d^4x\bigg[
  \overline{B}_v GB_v
 -\overline{T}^\mu_vQ_{\mu\nu}T^\nu_v
\nonumber \\ &&
 +{\cal C}\Big( \overline{T}^\mu_v\Theta_{\mu\nu}A^\nu_\Omega B_v
               +\overline{B}_v A^\mu_\Omega\Theta_{\mu\nu}T^\nu_v
          \Big)
\nonumber \\ &&
 +\frac{1}{2M}\Big[
     \overline{B}_v H^2B_v
    +\overline{T}^\mu_v R_{\mu\alpha}\widetilde{\cal Q}^{\alpha\beta}
                        R_{\beta\nu}T^\nu_v
\nonumber \\ &&\quad\quad~~
     -{\cal C}^2\Big(
      \overline{B}_v A^\mu_\Omega\Theta_{\mu\alpha}
                    \widetilde{\cal Q}^{\alpha\beta}
                    \Theta_{\beta\nu}A^\nu_\Omega B_v
     +\overline{T}^\mu_v\Theta_{\mu\alpha}A^\alpha_\Omega
               A^\beta_\Omega\Theta_{\beta\nu}T^\nu_v
                    \Big) \Big]
\nonumber \\ &&
     +O\bigg(\frac{1}{M^2}\bigg) \bigg] \bigg\}\,,
\label{Zeff}
\end{eqnarray}
where the determinants of the operators ${\cal G}$ and
${\cal Q}_{\mu\nu}$ are included in the normalization factor
${\cal N}$, and the terms containing external sources are omitted for the
sake of simplicity.

    Combining various relations, the leading pieces in the
$1/M$ expansion of the effective lagrangian
$$
  {\cal L}_v = {\cal L}^{(0)}_v(1/M^0)+{\cal L}^{(1)}_v(1/M^0)
               +O(1/M^2)
$$
in terms of the light components of baryon fields can be presented as
\begin{eqnarray}
{\cal L}^{(0)}_v &=& \overline{B}_v\Big(
  i(v\cdot D)-\gamma_5{{\,\not\!}{\hat A}}_\Omega \Big)B_v
\nonumber \\ &&
 -\overline{T}^\mu_v\bigg[\,
       i{\not\! v}(v\cdot D)g_{\mu\nu}
      -{1\over4}\gamma_\mu \gamma_\lambda i{\not\! v}(v\cdot D)
                \gamma^\lambda \gamma_\nu
\nonumber \\ &&\quad~~~~
      +{\cal H}\gamma_5\Big( {{\,\not}\! A}_\Omega g_{\mu\nu}
      +g_1\big( \gamma_\mu A_{\Omega\nu}
                      +A_{\Omega\mu}\gamma_\nu \big)
      +g_2\gamma_\mu {{\,\not}\! A}_\Omega \gamma_\nu\Big)
                  \bigg]T^\nu_v
\nonumber \\ &&
 +{\cal C}\Big(\,\overline{T}^\mu_v
               \big(g_{\mu\nu}+z\gamma_\mu\gamma_\nu\big)
               A^\nu_\Omega B_v
              +\overline{B}A^\mu_\Omega
               \big(g_{\mu\nu}+z\gamma_\mu\gamma_\nu\big)
               T^\nu_v
          \Big)\,.
\label{Lv0}
\end{eqnarray}

    The corrections of $O(1/M)$ have the following form:
\begin{eqnarray}
{\cal L}^{(1)}_v &=& \frac{1}{2M}\overline{B}_v\bigg\{
  -D^2+(v\cdot D)^2
  -\bigg(1-\frac{2}{3}{\cal C}^2\big(1-z-2z^2\big)\bigg)
   \big({\hat A}_\Omega^\mu\big)^2
\nonumber \\ &&\quad\quad\quad
 +\frac{i}{2}\sigma_{\mu\nu}\bigg[\,\big[D^\mu ,D^\nu\big]
 +2v^\mu \big[D^\nu ,(v\cdot D)\big]
\nonumber \\ &&\quad\quad\quad\quad\quad\quad
 +\bigg(1-\frac{1}{3}{\cal C}^2\big(1+2z+4z^2\big)\bigg)
  \big[{\hat A}_\Omega^\mu ,{\hat A}_\Omega^\nu\big]\bigg]
\nonumber \\ &&\quad\quad\quad
 +i\gamma_5 \Big( \big[D_\mu^\perp ,{\hat A}_\Omega^\mu\big]
 -i\sigma_{\mu\nu} \big\{D^{\perp\mu},{\hat A}_\Omega^\nu\big\}
            \Big)\bigg\}B_v
\nonumber \\
 &-&\frac{1}{2M}\overline{T}^\mu_v\bigg\{\,
  \frac{1}{3} g_{\mu\nu} \big(-D^2+(v\cdot D)^2\big)
 -\frac{i}{3}\sigma_{\mu\nu}D^2
\nonumber \\ &&\quad\quad
 +\frac{i}{6}\big( 2\sigma_{\alpha\beta}g_{\mu\nu}
                  +\gamma_\mu \sigma_{\alpha\beta}\gamma_\nu\big)
             \Big( \big[D^\alpha ,D^\beta\big]
                  +2v^\alpha \big[D^\beta ,(v\cdot D)\big]\Big)
\nonumber \\ &&\quad\quad
 -\frac{i}{3}\Big( {{\,\not}\! D}\sigma_{\mu\nu}{{\,\not}\! D}
         -(v\cdot D)\big[\sigma_{\mu\nu},{{\,\not}\! D}\big]
                     \Big)
 +\frac{1}{6}\big\{{{\,\not}\! D},
                   \gamma_\mu{{\,\not}\! D}\gamma_\nu\big\}
 +\frac{1}{3}(v\cdot D)^2 \gamma_\mu{\not\! v}\gamma_\nu
\nonumber \\ &&\quad\quad
 +\frac{1}{6}\big\{ (v\cdot D),
                   \gamma_\mu{{\,\not}\! D}\gamma_\nu\big\}
 +\frac{1}{6}\Big( (v\cdot D)\gamma_\mu{\not\! v}\gamma_\nu
                             {{\,\not}\! D}
                  +{{\,\not}\! D}\gamma_\mu{\not\! v}\gamma_\nu
                   (v\cdot D) \Big)
\nonumber \\ &&\quad\quad
 -i{\cal H}\gamma_5\bigg[\,
    \frac{1}{6}\Big( \big(4-g_1-4g_2\big)g_{\mu\nu}
                    +i\big(g_1+4g_2\big)\sigma_{\mu\nu}\Big)
    \big[ D^\perp_\alpha , A^\alpha_\Omega \big]
\nonumber \\ &&\quad\quad
 -\frac{i}{6}\Big( 4\sigma^{\alpha\beta}g_{\mu\nu}
                  -\big(g_1+4g_2\big)
                   \gamma_\mu\sigma^{\alpha\beta}\gamma_\nu\Big)
             \big\{ D^\perp_\alpha , A_{\Omega\beta}\big\}
\nonumber \\ &&\quad\quad
 +\frac{1}{6}\big[{{\,\not}\! A}_\Omega,
             \gamma_\mu{{\,\not}\! D}^\perp\gamma_\nu \big]
 +\frac{1}{3}\big(g_1+g_2\big)
             \big[\gamma_\mu{{\,\not}\! A}_\Omega\gamma_\nu,
             {{\,\not}\! D}^\perp \big]
\nonumber \\ &&\quad\quad
 -\frac{1}{3}g_1\big({{\,\not}\! D}^\perp\gamma_\mu A_{\Omega\nu}
            -A_{\Omega\mu}\gamma_\nu{{\,\not}\! D}^\perp\big)
 -\frac{5}{3}g_1\big(\gamma_\mu{{\,\not}\! D}^\perp A_{\Omega\nu}
            -A_{\Omega\mu}{{\,\not}\! D}^\perp\gamma_\nu\big)
                   \bigg]
\nonumber \\ &&\quad\quad
 -\bigg( \frac{2}{3}{\cal H}^2 \Big(1+(g_1+g_2)(g_1-2g_2)
        \Big)+{\cal C}^2 z^2\bigg)
        (A^\alpha_\Omega)^2 g_{\mu\nu}
\nonumber \\ &&\quad\quad
 +\bigg( \frac{2}{3}{\cal H}^2(g_1\!+g_2)
         (g_1\!-2g_2)\!-{\cal C}^2 z^2\bigg)
         i\sigma_{\mu\nu}(A^\alpha_\Omega)^2
 +\!\bigg( \frac{4}{3}{\cal H}^2 g^2_1\!+{\cal C}^2\bigg)
  A_{\Omega\mu}A_{\Omega\nu}
\nonumber \\ &&\quad\quad
 -\bigg( \frac{1}{3}{\cal H}^2 g_1 \big(3-g_1-4g_2\big)
        -{\cal C}^2 z\bigg)
  \Big( A_{\Omega\mu}{{\,\not}\! A}_\Omega\gamma_\nu
       +\gamma_\mu{{\,\not}\! A}_\Omega A_{\Omega\nu}\Big)
\nonumber \\ &&\quad\quad
 +{\cal H}^2\bigg[\,
  \frac{1}{6}(3+g_1^2+4g_2^2+2g_1g_2)g_{\mu\nu}
  i\sigma_{\alpha\beta}[A^\alpha_\Omega ,A^\beta_\Omega ]
\nonumber \\ &&\quad\quad
 +\frac{1}{6}(g_1^2+4g_2^2+2g_1g_2))\sigma_{\mu\nu}
  \sigma_{\alpha\beta}[A^\alpha_\Omega ,A^\beta_\Omega ]
 -\frac{i}{3}\gamma_\alpha\sigma_{\mu\nu}\gamma_\beta
  A^\alpha_\Omega A^\beta_\Omega
\nonumber \\ &&\quad\quad
 +\frac{1}{3}\big( g_1+g_2\big)
  \big\{ {{\,\not}\! A}_\Omega ,
         \gamma_\mu{{\,\not}\! A}_\Omega\gamma_\nu\big\}
 +\frac{1}{3} g_1
  \Big( {{\,\not}\! A}_\Omega\gamma_\mu A_{\Omega\nu}
       +A_{\Omega\mu} \gamma_\nu {{\,\not}\! A}_\Omega
  \Big)\bigg]
\nonumber \\ &&\quad\quad
 -\frac{1}{2}{\cal C}^2 z^2
  \gamma_\mu i\sigma_{\alpha\beta}\gamma_\nu
  [A^\alpha_\Omega ,A^\beta_\Omega ]
\bigg\}T^\nu_v\,.
\label{LvM}
\end{eqnarray}
    The lagrangian (\ref{LvM}) does not contain corrections for
transitions between spin-$1\over2$ and spin-$3\over2$ baryons which
appear at $O(1/M^2)$.

    Now we discuss the comparison of the results given by Eqs.\
(\ref{Lv0}) and (\ref{LvM}) with the effective lagrangians for light
velocity components of baryon fields derived on bases of the effective
meson-baryon lagrangian of Ref.\ \cite{Jenkins} which corresponds to
the Eq.\ (\ref{Ltot}) in absence of the off-shell terms contaning
gamma-matreces contracted with spin-${3\over2}$:
$$
{\cal L}_{tot} = \overline{B}\big(
  i{{\,\not}\! D}-M-\gamma_5{{\,\not\!}{\hat A}}_\Omega \big)B
 -\overline{T}^\mu\big(\,i{{\,\not}\! D}-M
 +{\cal H}\gamma_5{{\,\not}\! A}_\Omega\big)T_\nu
 +{\cal C}\big(\, \overline{T}^\mu A_{\Omega\mu} B
                 +\overline{B}A_{\Omega\mu}T^\mu \big)\,.
$$
   Following the path integral way described above we obtain in
this case the lowest order lagrangian in the form which coincides
with Eq.\ (\ref{Lv0}) if the off-shell terms are dropped out from
it.
    The corresponding corrections at $O(1/M)$ get the form
\begin{eqnarray}
{\widetilde {\cal L}}^{(1)}_v &=& \frac{1}{2M}\overline{B}_v\bigg\{
  -D^2+(v\cdot D)^2
  -(1-{\cal C}^2)\big({\hat A}_\Omega^\mu\big)^2
\nonumber \\ &&\quad\quad\quad
 +\frac{i}{2}\sigma_{\mu\nu}\Big( \big[D^\mu ,D^\nu\big]
 +2v^\mu \big[D^\nu ,(v\cdot D)\big]
 +\big[{\hat A}_\Omega^\mu ,{\hat A}_\Omega^\nu\big]\Big)
\nonumber \\ &&\quad\quad\quad
 +i\gamma_5 \Big( \big[D_\mu^\perp ,{\hat A}_\Omega^\mu\big]
 -i\sigma_{\mu\nu} \big\{D^{\perp\mu},{\hat A}_\Omega^\nu\big\}
            \Big)\bigg\}B_v
\nonumber \\ &&
 -\overline{T}^\mu_v\bigg\{\,
  g_{\mu\nu}\bigg[ -D^2+(v\cdot D)^2
                   -{\cal H}^2 \big( A_\Omega^\mu\big)^2
\nonumber \\ &&\quad\quad
 +\frac{i}{2}\sigma_{\alpha\beta}\Big(
             \big[D^\alpha ,D^\beta\big]
            +2v^\alpha \big[D^\beta ,(v\cdot D)\big]
            +{\cal H}^2\big[A_\Omega^\alpha ,A_\Omega^\beta\big]
                                 \Big)
\nonumber \\ &&\quad\quad
 -i{\cal H}\gamma_5 \Big(\big[D_\alpha^\perp ,A_\Omega^\alpha\big]
 -i\sigma_{\mu\nu}\big\{D^{\perp\alpha}, A_\Omega^\beta\big\}
                    \Big)\bigg]
+{\cal C}^2 A_{\Omega\mu} A_{\Omega\nu} \bigg\}T^\nu_v\,.
\label{LvMprime}
\end{eqnarray}
    The comparison of Eqs.\ (\ref{LvM}) and (\ref{LvMprime}) shows
that the taking into account the off-shell terms in the effective
lagrangian (\ref{Ltot}) leads not only to the additional off-shell
terms in the expressions for $1/M$-correction, but also modifies
the terms which does not disappear on mass shell of spin-${3\over2}$.

    As the expressions derived above for ${\cal L}^{(0)}_v$ and
${\cal L}^{(1)}_v$ are faily cumbersome, it is tempting to look
for a simpler representation of the heavy and light components
of the spin-${3\over2}$ field.
    An interesting alternative was pursued in Ref.\ \cite{Kambor}
based on the definitions
\begin{eqnarray}
&&
T^\mu = \mbox{e}^{-iM(v\cdot x)}(T^\mu_v +t^\mu_v)\,,\quad
\nonumber\\ &&
T^\mu_v = \frac{1+{\not\! v}}{2}  P^{(3/2)\mu\nu}_v T_\nu\,,\quad
\nonumber\\ &&
t^\mu_v = \bigg( g^{\mu\nu}
                -\frac{1+{\not\! v}}{2}P^{(3/2)\mu\nu}_v\bigg)T_\nu
        =\bigg( \frac{1-{\not\! v}}{2}P^{(3/2)\mu\nu}_v
               +P^{(1/2)\mu\nu}_v\bigg) T_\nu\,,
\label{Tvnew}
\end{eqnarray}
using spin projection operators \cite{Aurilia}
\begin{eqnarray*}
&&
P^{(3/2)\mu\nu}_v =
   g^{\mu\nu} -\frac{1}{3}\gamma_\mu \gamma_\nu
  -\frac{2}{3}v_\mu v_\nu
  -\frac{1}{3}{\not\! v}\big(\gamma_\mu v_\nu-v_\mu\gamma_\nu\big)\,,
\\ &&
P^{(1/2)\mu\nu}_v = g_{\mu\nu}-P^{(1/2)\mu\nu}_v\,.
\end{eqnarray*}
    With this definition the light component $T^\mu_v$ corresponds to
a massless pure spin-${3\over2}$ state, which satisfies the constraints
\begin{equation}
  v_\mu T^\mu_v =0\,,\quad \gamma^\nu T^\mu_v =0\,,
\label{propnew}
\end{equation}
while the heavy component $t^\mu_v$ is the mixture of spin-${1\over2}$
and spin-${3\over2}$ states.
    In this case the operators $Q_{\mu\nu}$, ${\cal Q}_{\mu\nu}$
and $R_{\mu\nu}$ in Eq.\ (\ref{Ltotv}) get the form
\begin{eqnarray*}
&&
  Q_{\mu\nu} =\, \big[ i(v\cdot D)
      +{\cal H}\gamma_5{{\,\not}\! A}_\Omega \big] g_{\mu\nu}
\nonumber \\ &&
  {\cal Q}_{\mu\nu} = \big[i{{\,\not}\! D}
                           -(1-{\not\! v})M\big]g_{\mu\nu}
      -{1\over4}\gamma_\mu \gamma_\lambda
       \big[ i{{\,\not}\! D}-(1-{\not\! v})M\big]
       \gamma^\lambda\gamma_\nu
      +{\cal H}\gamma_5 \Omega_{\mu\nu}\,,
\nonumber \\ &&
  R_{\mu\nu} =\, i{{\,\not}\! D}g_{\mu\nu}
      +{\cal H}\gamma_5\big( {{\,\not}\! A}_\Omega g_{\mu\nu}
      +g_1 A_{\Omega\mu}\gamma_\nu \big)\,.
\end{eqnarray*}
    Due to the simple form of the operator $Q_{\mu\nu}$ and the
constraints (\ref{propnew}), the definitions (\ref{Tvnew}) seem to be
more preferable than we have used above.
    Unfortunately, it can be shown that the inverse operator
${\cal Q}^{-1}_{\mu\nu}$ does not exist in this case and the procedure
of integrating out the heavy component $t^\mu_v$ can not be performed.

   This work was supported in part by the Kernforschungszentrum
J\"ulich under contract No. ER-41154523.


\end{document}